# Enhancement of nonvolatile polarization and pyroelectric sensitivity in Lithium tantalate (LT)/ Poly (vinylidene fluoride) (PVDF) nano composite

**S. Satapathy[*1], P. Kumar. Gupta[1] and K. BabiRaju. Varma[2]**

[1]Laser Materials Development & Device Division, Raja Ramanna Centre for Advanced Technology, Indore 452013, India

[2]Materials Research Centre, Indian Institute of Science, Bangalore 560 012, India

*Corresponding Author:

Tel: 91 0731 2488660

Fax: 91 0731 2488650

Email: [a] srinu73@cat.ernet.in

## Abstract

For pyroelectric sensor application, materials having large ferroelectric polarization at low applied field and high pyroelectric sensitivity are required. PVDF is one of the ferroelectric polymers, which is being used for pyroelectric detector application. Following properties of PVDF such as low density, low acoustic impedance, wide bandwidth, flexibility, toughness, and ease of fabrication into complex patterns and arrays make it suitable for large area detector applications. But the PVDF has low value of pyroelectric coefficient ($p$). High field (above 1200kV/cm) is required to pole the PVDF film. To reduce poling field and to get high value of ferroelectric polarization at low poling field, nano composite films of LT/PVDF have been prepared. Since LT is ferroelectric (max. poling field ~ 200kV/cm and ($p$) ~ 2.3 * $10^{-8}$ C $cm^{-2}$/ K) in nature, the nonvolatile polarization of composite increased from 0.014$\mu$C/$cm^2$ to 2.06$\mu$C/$cm^2$ at an applied field of 150kV/cm as the volume fraction of LT ($f_{LT}$) nano particle increases from 0.0 to 0.17. Similarly the increase of $f_{LT}$ from 0.0 to 0.17 results in increase of pyroelectric voltage sensitivity from 3.93 V/J to 18.5V/J.

# 1. Introduction.

Nano composites of polymer have superior physical properties compared to pure polymer [1-6]. Materials having high pyroelectric sensitivity are required for pyroelectric sensor application. The material should be poled at low electric field and the poling field is more important in case of thin film.

Considerable attention has been devoted in the past decade to study piezoelectric and pyroelectric properties of PVDF because of its favorable properties as a transducer component material. The unique properties of PVDF transducers include low density, low acoustic impedance, wide bandwidth, flexibility, toughness, and ease of fabrication into complex patterns and arrays. These properties are ideally suited to yield a great variety of designs for sensors and actuators in many different fields of application at a potentially low cost [7]. High field (greater than 1200KV/cm) is required to pole the PVDF film [8]. The dielectric constant ($\varepsilon$) is 9 at 60Hz and the reported pyroelectric coefficient ($p$) is $0.27 \times 10^{-8}$ C cm$^{-2}$/ K [9]. For high figure merit of pyroelectric material, it should have high $p$, low $\varepsilon$ and low thermal conductivity (K) since these result in relatively large changes in voltage and temperature respectively [10].

So for better performance of detector, responsivity ($R_I$) must be as high as possible. And for higher value of $R_I$, the pyroelectric coefficient must be high. But The PVDF has low value of pyroelectric coefficient ($p$). The value of Coercive field (E$_c$) is very high to pole the PVDF film. Due to low value of ($p$) its sensitivity is also low. Its poor heat dissipation makes it highly vulnerable to thermal damage.

Lithium tantalate exhibits unique electro-optical, pyroelectric and piezoelectric properties combined with good mechanical and chemical stability and, wide transparency range and high optical threshold. It is rugged, non hygroscopic, and chemically stable. This makes $LiTaO_3$ well-suited for numerous applications including electro-optical modulators, pyroeletric detectors, optical waveguide and SAW substrates, piezoelectric transducers etc. With a Curie temperature above 600 $^0$C [11], it is resistant to depoling and exhibits linear response throughout a wide operating range [12]. The required field to pole the LT (congruent) is 200KV/cm and for stoichiometric composition of LT the required field is 17kV/cm [13]. The dielectric constant (ε) of LT nano particle is around 40 [14]. The pyroelectric coefficient of LT is 2.3 * $10^{-8}$ C $cm^{-2}$/ K [12]. The pyroelectric coefficient of LT is greater than PVDF.

For enhancing the value pyroelectric sensitivity of PVDF, there is need to make composite of PVDF with a filler material which having very high value of pyroelectric coefficient, low poling field and dielectric constant comparable to that of PVDF. So we choose lithium tantalate nanoparticles to increase the nonvolatile polarization at small applied filed and pyroelectric coefficient of PVDF. At application of low field upto 200kV/cm, there is almost no non volatile polarization arises in PVDF. Here we report the increase of ferroelectric polarization on application of low electric field and increase of pyroelectric voltage sensitivity in LT/PVDF nano composite film compared to pure PVDF film. As the volume fraction of LT ($f_{LT}$) increases from 0.0 to 0.17 in LT/PVDF nano composite the ferroelectric polarization increases from 0.014μC/$cm^2$ to 2.06μC/$cm^2$ at an applied field of 150kV/cm. Similarly the increase of $f_{LT}$ from 0.0 to0.17 results in increase of pyroelectric voltage sensitivity from 3.93 V/J to 18.5V/J.

## 2. Experiment.

Nano particles of LT were prepared by sol-gel process through alkoxide route. Spherical well separated nano particles (20-40nm) of LT were obtained by adding oleic acid (capping ligand) [14] to sol of LT. These nano particles were dispersed in PVDF solution which was prepared using Dimethyl sulphoxide (DMSO) solvent at $90^0$C. The LT/PVDF composites were cast on glass plates at same temperature. The free standing thin films of thickness 30μm were annealed at $90^0$C for 5 hours. The LT/PVDF composites films with different volume fractions of LT were prepared in order to study the change in ferroelectric polarization behavior with the volume fraction of the filler. The phases of composite were studied using X-ray diffraction. The surface of samples was examined by scanning electron microscopy (SEM) and atomic force microscopy (AFM). The dielectric measurements were carried out by coating the surface of the pure PVDF and composites with gold. The dielectric measurements were carried out using precision impedance analyzer (HP 4194A) in frequency range of 100Hz to $10^6$Hz. The ferroelectric hysteresis properties were studied using ferroelectric loop tracer (RT6000 of Radiant technologies) at 1Hz. The pyroelectric sensors were fabricated using PVDF and LT/PVDF composite films. The surface of sensor was coated with gold by thermal evaporation technique. Pyroelectric voltage sensitivities of composite films were obtained using pulsed Nd: YAG laser of 17ns pulse width.

## 3. Results and discussion.

The X-ray diffraction patterns confirm the proper phases of PVDF and LT nano particles embedded in composite. From SEM and AFM it is clear that in pure PVDF the crystalline regions are separated by non crystalline regions. In LT/PVDF composites nano particles (20-40nm) of LT are uniformly distributed in PVDF which is confirmed from SEM and AFM. For $f_{LT}$ =0.047, the connectivity of composite is 0-3. As the volume fraction of LT increases from 0.047 to 0.17 the nano particles are connected each other and start forming network along all direction. From dielectric measurement it is clear that the dielectric constant loss is almost constant as the volume fraction of LT increases in PVDF matrix in high frequency region ($10^3$Hz to $10^6$Hz). In this frequency the dielectric loss of composite is almost same as that of pure PVDF. So definitely there is increase of dipole moments due to addition of LT nano particles.

To see the changes in ferroelectric polarization in nano composites of LT/PVDF hysteresis and pulsed polarization positive up-negative down (PUND) characterization have been carried out. Figure 1(a) shows P-E hysteresis loop of pure PVDF film. The polarization of 0.17μC/cm$^2$ is obtained at an applied electric field of 400kV/cm at 1Hz. Figure 1(b) and 1(c) show the hysteresis loops of LT/ PVDF nano composites for $f_{LT}$ = 0.047 and 0.09 respectively. In case of $f_{LT}$ = 0.047, the polarization increases slightly to 0.23 μC/cm$^2$ at same applied field 400kV/cm. The polarization of composite further increases to 1.04μC/cm$^2$ as volume fraction of LT increases to 0.09. The dielectric loss increases as the volume fraction of LT increases in low frequency range which reflects in hysteresis loop at an applied filed 400kV/cm (figure 1c). So PUND measurements were carried out to get proper estimate of non volatile polarization of composite thin films (discussed in next section). But when the volume fraction of LT nano particle increases to 0.17, then there is a sharp increase of polarization compared to PVDF. At an applied field of 170kV/cm, the value of polarization is 2.6μC/cm$^2$ in case LT/PVDF ($f_{LT}$ = 0.17) composite ((figure 1(d)). From hysteresis plot it is clear that at this applied field (170kV/cm) the polarization of pure PVDF is almost negligible (0.05μC/cm$^2$). **To achieve** **polarization of 2.6μC/cm² in PVDF, one has to apply more than 1000kV/cm polarization.** So for practical device application film of LT/ PVDF ($f_{LT}$= 0.17) nano composite may replace pure PVDF.

To confirm the polarization arises due to intrinsic dipoles, pulsed polarization positive up-negative down (PUND) measurements of composites have been carried out with delay time of 1second. The PUND results are shown in fig. 2. One can see from fig. 2(a) that with increasing applied voltage, the nonvolatile polarization (dP = P*(switched polarization) –P^ (nonswitched polarization)) starts to arise at 100kV/cm and increases sharply up to 600kV/cm. The dP value reaches 0.25 μC/cm² at 400kV/cm which is less than 2P$_r$ value (0.34 μC/cm² at 400kV/cm) observed in hysteresis analysis. This result demonstrates that for pure PVDF the nonvolatile polarization value is 0.125μC/cm² at an applied field 400 kV/cm.

But for composite in which $f_{LT}$ is 0.047, the nonvolatile polarization starts to arise at 100kV/cm and increases sharply up to 400 kV/cm(figure 2(b)). The dP value reaches 0.43 μC/cm² at 400kV/cm which is consistent with 2P$_r$ value (0.46 μC/cm² at 400kV/cm). For $f_{LT}$ = 0.09, the nonvolatile polarization (dP = P*(switched polarization) –P^ (nonswitched polarization)) is 2.05 μC/cm² at 400kV/cm (fig. 2(c)) which is consistent with the 2P$_r$ values obtained from hysteresis (i. e. 2.08 μC/cm² at 400kV/cm for $f_{LT}$ = 0.09). Similarly for $f_{LT}$ = 0.17 the nonvolatile polarization is 4.13 μC/cm² at 150 kV/cm (figure 2(d)) which is also consistent with the 2P$_r$ values obtained from hysteresis (4.4μC/cm² at 150 kV/cm for $f_{LT}$ = 0.17). PUND measurement shows a polarization of 0.0142μC/cm² for pure PVDF at 150kV/cm (figure 2(a)). From PUND measurement it is clear that the intrinsic polarization is due to ferroelectricity not due to leakage. PUND measurement also shows the ferroelectric polarization increases drastically from 0.0142 $\mu C/cm^2$ for pure PVDF to 2.06 $\mu C/cm^2$ for LT/PVDF ($f_{LT}$ = 0.17) composite at an applied field 150kV/cm.

Here question may arise why we apply low field to PVDF and compare with LT/PVDF composite. The maximum coercive field of LT is 200kV/cm. When volume fraction of LT increases in PVDF application of high field (greater than 200kV/cm) is impossible due to the conduction path provided by nano particles. At high field (greater than 1200kV/cm) the PVDF may show high polarization but direct poling of thin PVDF film is always prone to damage of thin film. So after addition of LT nano particles (volume fraction 0.17), the poling field has been reduced and there is appreciable amount of nonvolatile polarization i. e. 2.06 $\mu C/cm^2$ generates which is comparable to saturation polarization of triglycine sulphate (2.9 $\mu C/cm^2$).

In order to verify the pyroelectric properties of LT/PVDF composite films, pyroeletric sensors, operating in laser energy meter mode, of these films were fabricated. The pyroelectric voltages generated at constant energy of laser pulse from these films were recorded. For energy meter application, the condition pulse width of laser (w) << thermal time constant of laser must be satisfied. [15] This indicates that the thickness of sensor much greater than the diffusion depth so that during the radiation pulse little heat will flow out of the rear surface of the ferroelectric. Then the output voltage of sensor can be expressed as

$$V_0(t) = \left( \frac{Ap}{c'd} \right) \left( \frac{1}{C} \right) \exp\left( \frac{-t}{\tau_e} \right) \int_0^\omega \exp\left( \frac{t}{\tau_e} \right) F(t)\, dt \qquad (1)$$

Where, $\tau_e$ = RC, the electrical time constant of the circuit and w is the pulse width of the source. The 'F(t)' is the space-averaged energy flux per unit area absorbed by the detector.

The 'p' is the pyroelectric coefficient of the material and A is the active area of the sensor. The $c^{/}$ is volume specific heat of the material and $d$ is thickness of sensor material.

If the electrical time constant is long compared with the duration of incident signal, and $\tau_e >> $ w then $eq^n.1$ simplifies to

$$V_0(t) = \left(\frac{p}{c^{/}d}\right)\left(\frac{1}{C}\right)e^{-t/\tau_r}\left(A\int_0^w F(t)dt\right) \qquad (2)$$

where $\left(A\int_0^w F(t)dt\right)$ = E (total energy of the pulse fall on the detector)

So out put voltage at any time for the condition w<< $\tau_e$<< $\tau_T$ is

$$V_0(t) = \left(\frac{p}{c^{/}d}\right)\left(\frac{1}{C}\right)e^{-t/\tau_r}E \qquad (3)$$

If input energy of the pulse falls on the coating material (coated on pyroelectric material), then equation (3) can be written as

$$V_0(t) = \left(\frac{p}{c^{/}d}\right)\left(\frac{1}{C}\right)e^{-t/\tau_r}\eta E \qquad (4)$$

where $\eta$ is emissivity of the coating material.

Figure 3(a) shows the output signal generated from LT/PVDF films (for $f_{LT}$ = 0.0, 0.047, 0.09 and 0.17) when Nd: YAG laser pulse of width 17ns of energy 2.6mJ falls on these sensors. The magnitude of generated output voltage at constant energy 2.6mJ increases from 11.31mV to 49mV as $f_{LT}$ increases from 0 to 0.17. Figure 3(b) shows the output voltage response of composite films with variation of input energy. As input energy of laser pulse increases the output voltage generated from each film increases linearly. So from figure 3(b) it is clear that the pyroelectric voltage sensitivity of PVDF increases with addition of LT nano particles. The sensitivity increases from 3.9V/J to 18.5V/J as the volume fraction of LT increases from 0.0 to 0.17. The sensitivity of LT/PVDF composite ($f_{LT}$ = 0.17) has increased 4.7 times in comparison to pure PVDF.

The pyroelectric coefficients of PVDF and LT/PVDF composites have been calculated using equation (4). The pyroelectric coefficient increases from 0.128 $\times 10^{-8}$ C/cm$^2$ K to 1.47$\times 10^{-8}$C/cm$^2$ K as volume fraction of LT increases from 0.0 to 0.17.

The ferroelectric polarization and pyroelectric voltage sensitivity of LT/PVDF composite increases compared to PVDF as the volume fraction of LT increases in PVDF. Why ferroelectric polarization of LT/PVDF composite film is increased compared to pure PVDF?

Ferroelectric nano particles smaller than the ferroelectric domain size have a single domain structure [16, 17]. The characteristic ferroelectric to paraelectric transition temperature i. e. Curie temperature decreases as the particle size of ferroelectric material decreases. In case of nano particles, there is no long range ferroelectric order. The energy barrier separating alternative polarization states of the nano particle decreases as the size decreases. The ferroelectric polarization value increases if the cooperative interaction between nano particles increases. The random distribution of nano particles in a matrix results in random distribution of polarization directions, thus the system in a super paraelectric state below Curie temperature. For LT nano particle of size 20-40nm, the Curie temperature decreases from 660 to 492$^0$C [14]. When volume fraction of LT in PVDF matrix is 0.047, then the distribution of nano agglomeration is random, results in very less increase of nonvolatile polarization. As volume fraction of LT increases in PVDF matrix the cooperative interaction between nano particle increases. We have already described for volume fraction of 0.17, the connectivity of nanoparticles increases along surface of the PVDF film as well as along the ferroelectric measurement direction which in turn increase cooperative interaction between nano particles.

The coercive field of LT is 200kV/cm. The spontaneous polarization of LT is $60\mu C/cm^2$ at an applied filed 200kV/cm. But high filed (1200kV/cm) is required to pole the PVDF film. So at small applied filed, the LT nano particles easily poled in composite in comparison to PVDF crystalline regions. So as LT volume fraction increases more nano particles poled at application of 150kV/cm. So the nonvolatile polarization increases as the volume fraction of LT increases. We observed large increase of ferroelectric polarization i.e. from $0.014\mu C/cm^2$ for pure PVDF to $2.06\mu C/cm^2$ for LT/PVDF ($f_{LT} = 0.17$) at an applied field of 150kV/cm.

## 4. Conclusion

The results show number of important properties of LT/PVDF nano composite.

1. From hysteresis and PUND measurement it is clear that the ferroelectric polarization of PVDF increases with addition LT nano particles at low poling field. As the volume fraction of LT increases in LT/PVDF composite, the ferroelectric polarization increases. The ferroelectric polarization increases from $0.014\mu C/cm^2$ for pure PVDF to $2.06\mu C/cm^2$ for LT/PVDF ($f_{LT} = 0.17$) at an applied field of 150kV/cm. So there is almost 140 times increase of nonvolatile polarization in LT/PVDF composite at an applied filed 150kV/cm compared to pure PVDF.

2. Applying low field 150kV/cm one can pole LT/PVDF composite film.

3. The pyroelectric coefficient increases from $0.128 \times 10^{-8} C/cm^2$ K to $1.47 \times 10^{-8} C/cm^2$ K as volume fraction of LT increases from 0.0 to 0.17. The pyroelectric voltage sensitivity of PVDF also increases with addition of LT nano particles. As the volume fraction of LT increases in LT/PVDF composite from 0.0 to 0.17, the pyroelectric voltage sensitivity increases from 3.9V/J to 18.5V/J. There is almost 4.7 times increase of sensitivity in LT/PVDF composite ($f_{LT}$ = 0.17) compared to pure PVDF.

4. The report also shows the potential use of LT oxide nano particles. Growth of LT single crystal is very difficult. Similarly to pole PVDF large field required. By use of nano particles in PVDF one can decrease poling field of PVDF appreciably and get high polarization. So LT/PVDF nano composite can be used as an alternative to single crystal $LiTaO_3$ and pure PVDF for pyroelectric sensor application.

5. Potential use of nanocomposite of LT/PVDF as pyroelectric laser energy meter has been demonstrated.

**Acknowledgement** The authors are grateful to Dr. S. M. Gupta for his help in characterization.

**Figure Caption:**

Figure 1. P-E hysteresis loop of (a) pure PVDF and of LT/PVDF composites for $f_{LT}$ = (b) 0.047 (c) 0.09 and (d) 0.17.

Figure 2. PUND measurements for varying electric fields of (a) pure PVDF and of LT/PVDF composites for $f_{LT}$ = (b) 0.047 (c) 0.09 and (d) 0.17.

Figure 3. (a) Pyroelectric signals generated from LT/PVDF composites for $f_{LT}$ = 0.0, 0.047, 0.09 and 0.17, when Nd: YAG laser pulse of energy 2.6mJ falls on the sensors. (b) Voltage sensitivity curve of LT/PVDF composites for $f_{LT}$ = 0.0, 0.047, 0.09 and 0.17.

**Figure 1**

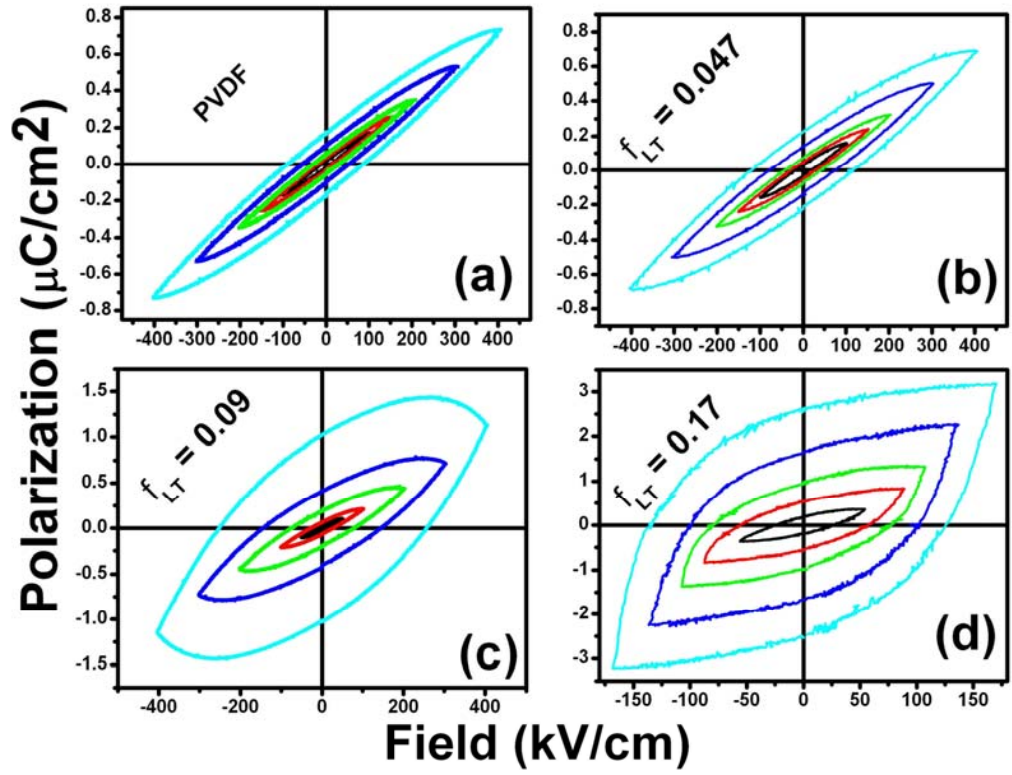

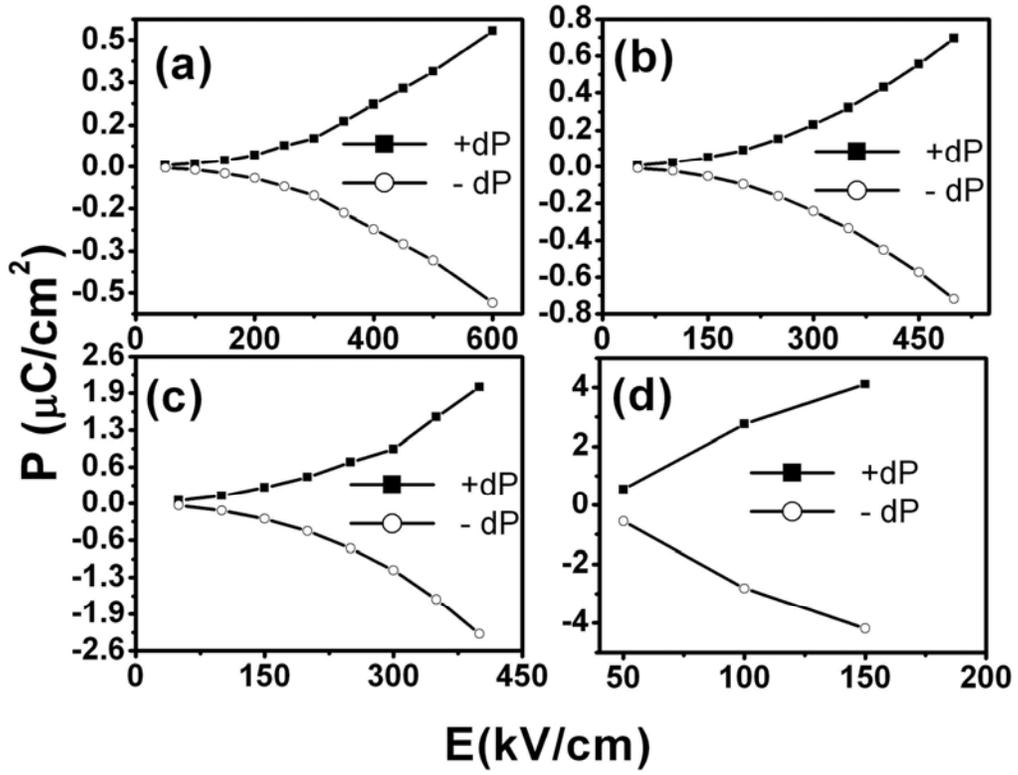

**Figure 3**

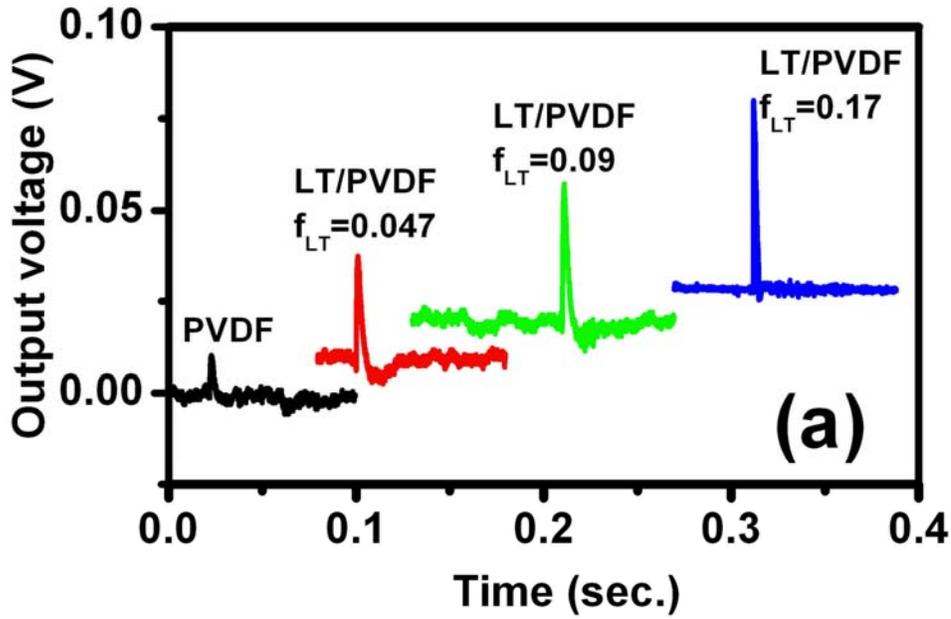

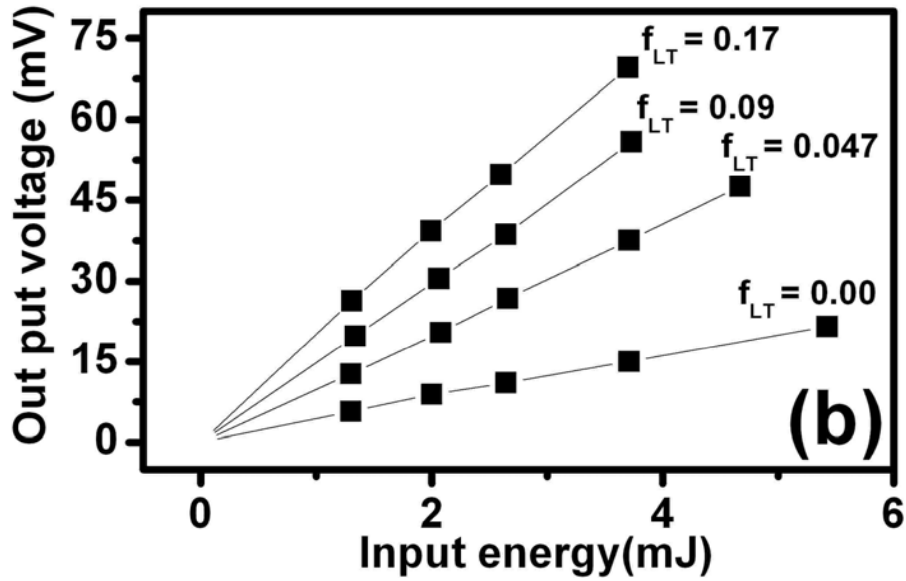